\documentclass[11pt,floatfix,notitlepage,byrevtex]{article}
\usepackage{graphicx,epstopdf}
\usepackage[totalwidth=500pt, totalheight=650pt]{geometry}
\usepackage{pstricks,pst-plot,pst-grad,pst-3dplot,pstricks-add,pst-node}
\usepackage{mathrsfs}   
\usepackage{amsmath}    
\usepackage{amssymb}    
\usepackage{CJKutf8,CJKnumb}
\usepackage{hyperref}
\usepackage{CJK}
\usepackage{amsfonts}
\usepackage{caption2,graphics,placeins,verbatim}
\usepackage{diagbox}
\usepackage[numbers,sort&compress]{natbib}
\usepackage{authblk}

\allowdisplaybreaks

\title{Euclidean (A)dS spaces over $p$-adic numbers}
\author{Feng Qu\thanks{qufeng@syu.edu.cn}}
\affil{\small{The Normal College, Shenyang University, Shenyang, P.~R.~China}}
\date{}

\begin{document}
\bibliographystyle{unsrt}
\maketitle

\begin{abstract}
With the help of Wick rotation over $p$-adic numbers $\mathbb{Q}_p$, the $p$-adic version of Euclidean $\textrm{dS}_2$ space(noted as $p\textrm{dS}_2$) is obtained based on $p\textrm{AdS}_2$($p$-adic version of Euclidean $\textrm{AdS}_2$ space), the latter of which is already known. The corresponding embedding equations are also found. The distances $D(X,Y)$'s on $p\textrm{(A)dS}_1$ and $p\textrm{AdS}_2$ have intuitive explanations. On the graph representations of $\mathbb{Q}_p$ and $\mathbb{Q}_{p^2}$, namely Bruhat-Tits trees $\textrm{T}_{p}$ and $\textrm{T}_{p^2}$, $D(X,Y)$ is found to be the inverse of distance between a particular subgraph and the line connecting $X$ and $Y$.
\end{abstract}

\section{Introduction}

The general covariance principle claims that physical laws are invariant under the change of coordinates. In~\cite{Volovich:1987wu}, a similar principle(number field invariance principle) is proposed which claims that  physical laws should be invariant under the change of number fields. It means physical laws are the same no matter what number field is used by the observer. Such number field should include rational numbers $\mathbb{Q}$ since all measurement results are written with them. To verify this ``number field invariance principle'', it is necessary to study physics over other number fields besides real numbers $\mathbb{R}$. Studying physics over $p$-adic numbers $\mathbb{Q}_p$ is one example, such as~\cite{Volovich_1987,Freund:1987kt,FREUND1987191,Zabrodin:1988ep,Vladimirov:1988bd,Smirnov:1992sr,Gubser:2017vgc}. The properties of $\mathbb{Q}_p$ can be found in~\cite{Vladimirov:1994wi}. This paper is devoted to the investigation of spaces over $\mathbb{Q}_p$. The $p$-adic version of Euclidean $\textrm{AdS}_2$ is proposed in~\cite{Heydeman:2016ldy,Gubser:2016guj}, and it is widely used when combining with the anti-de Sitter/conformal field theory correspondence~\cite{Maldacena:1997re,Gubser:1998bc,Witten:1998qj,Gubser:2016htz,Bhattacharyya:2017aly,Qu:2018ned,Gubser:2018cha,Ebert:2019src,Huang:2019nog,Huang:2020qik}. Considering that de Sitter and anti-de Sitter are two important spaces over $\mathbb{R}$ whose Euclidean versions are sphere and hyperboloid in high-dimensional spaces, we study the $p$-adic version of Euclidean (A)dS spaces(noted as $p\textrm{(A)dS}$) including the embedding of $p\textrm{(A)dS}_2$ and the analysis of their subspaces $p\textrm{(A)dS}_1$. There are also some other papers studying the embedding problem, such as~\cite{Guilloux_2016,Bhowmick:2018bmn}.

This paper is organized as follows. Section~\ref{sec:eadsoverRandqp} includes a review of Euclidean $\textrm{(A)dS}_2$ spaces(noted as $\textrm{E(A)dS}_2$) over $\mathbb{R}$ and provides some basic knowledge of $\mathbb{Q}_p$. At the beginning of section~\ref{sec:disonbdy}, we define distances on the boundaries of $\textrm{T}_p$ and $\textrm{T}_{p^2}$ which are graph representations of $\mathbb{Q}_p$ and its two-dimensional unramified extension $\mathbb{Q}_{p^2}$~\cite{Gubser:2016guj}. This distance has a very intuitive explanation on the graph: it is inversely proportional to the distance(defined at the beginning of section~\ref{sec:disonbdy}) between the line connecting these two points and a selected subgraph(the reference subgraph). It is found at the end of section~\ref{sec:disonbdy} that $p\textrm{AdS}_2$'s reference subgraph is one subgraph $\textrm{T}_p$ of $\textrm{T}_{p^2}$.  In section~\ref{sec:peads}, firstly we clarify  the Wick rotation over $\mathbb{Q}_p$, which is actually noticed in some papers such as~\cite{Stoica:2018zmi}. Secondly, we find the embeddings of  $p\textrm{(A)dS}_2$. Thirdly, it is shown that $p\textrm{dS}_1$ and $p\textrm{AdS}_1$'s reference subgraphs are one vertex and one line of $\textrm{T}_p$ respectively. The last section~\ref{sec:sumanddis} is the summary and discussion.

\section{$\textrm{E(A)dS}_2$ and basic knowledge of $\mathbb{Q}_p$}\label{sec:eadsoverRandqp}

$\textrm{E(A)dS}_2$ can be defined as hypersurfaces in $\mathbb{R}^3$ and $\mathbb{R}^{2,1}$
\begin{gather}
\mathbb{R}^3:~ds^2=dX_0^2+dX_1^2+dX_2^2~,~\mathbb{R}^{2,1}:~ds^2=dX_0^2+dX_1^2-dX_2^2~,
\\
\textrm{E(A)dS}_2:~X_0^2+X_1^2\pm X_2^2=\pm L^2~,
\end{gather}
where ``$\pm$'' should be replaced by ``$+$'' for $\textrm{EdS}_2$, and ``$-$'' for $\textrm{EAdS}_2$. With proper coordinate transformations, metrics of $\textrm{E(A)dS}_2$ can be rewritten as
\begin{gather}
\textrm{E(A)dS}_2:~ds^2=\frac{4L^2}{(1\pm x_0^2\pm x_1^2)^2}(dx_0^2+dx_1^2)~,
\\
\left\{
\begin{aligned}\label{trans1}
&x_0=\frac{X_0}{L+X_2}~,~x_1=\frac{X_1}{L+X_2}~,
\\
&X_0=\frac{2x_0L}{1\pm x_0^2\pm x_1^2}~,~X_1=\frac{2x_1L}{1\pm x_0^2\pm x_1^2}~,X_2=\frac{1\mp x_0^2\mp x_1^2}{1\pm x_0^2\pm x_1^2}L~.
\end{aligned}
\right.
\end{gather}
These coordinate transformations are actually perspective projections with the center of projection $(X_0=0,X_1=0,X_2=-L)$ and the projection plane $X_2=0$. Two $(x_0,x_1)$ coordinate systems(one of $\textrm{EdS}_2$ and one of $\textrm{EAdS}_2$) are global coordinate systems if ignoring the center of projection and not imposing additional constraints such as $x_0^2+x_1^2<1$ for $\textrm{EAdS}_2$.

There is another useful coordinate system for $\textrm{EAdS}_2$
\begin{gather}
\textrm{EAdS}_2:~ds^2=\frac{L^2}{x_0'^2}(dx_0'^2+dx_1'^2)~,
\\
\left\{
\begin{aligned}\label{trans2}
&x_0'=\frac{L}{X_2-X_0}~,~x_1'=\frac{X_1}{X_2-X_0}~,
\\
&X_0=\frac{x_0'^2+x_1'^2-1}{2x_0'}L~,~X_1=\frac{x_1'L}{x_0'}~,~X_2=\frac{x_0'^2+x_1'^2+1}{2x'_0}L~.
\end{aligned}
\right.
\end{gather}
It is usually used to describe one half of the entire $\textrm{EAdS}_2$ which is achieved by demanding that $x_0'>0$ or $x_0'<0$. While it is still a global coordinate system if not imposing additional constraints.

\begin{figure}
	\centering
	\includegraphics[width=0.8\textwidth]{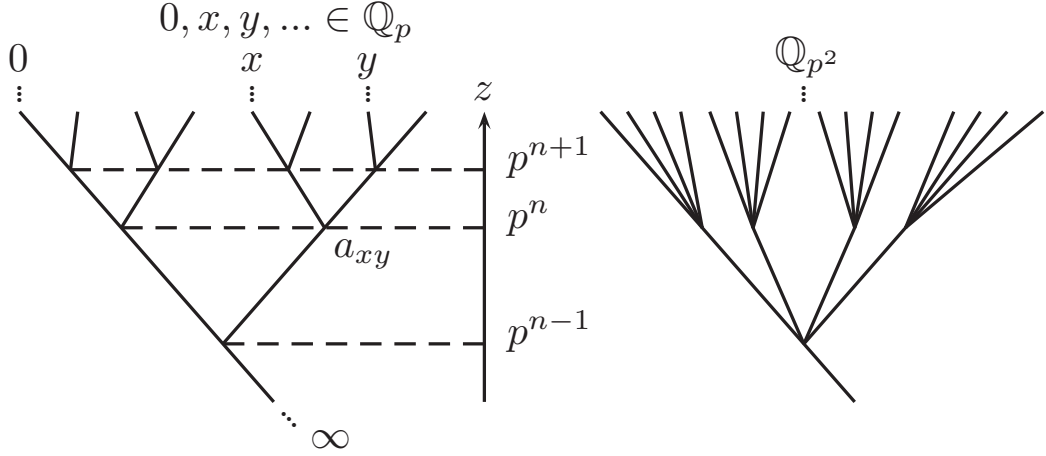}
	\caption{\label{tpandtp2}Taking $p=2$ as an example, $\textrm{T}_p$ and $\textrm{T}_{p^2}$ are trees(graphs with no loop) whose upper boundaries are $\mathbb{Q}_p$ and $\mathbb{Q}_{p^2}$ respectively.}
\end{figure}

$\mathbb{Q}_p$ can be regarded as the boundary of Bruhat–Tits tree($\textrm{T}_p$) which is a graph without loops, and each vertex has $p+1$ nearest neighboring vertices. Refer to the left in Fig~\ref{tpandtp2}. 
There is a $z$ coordinate satisfying
\begin{gather}
|x-y|_p=|z(a_{xy})|_p=|p^n|_p=p^{-n}~,~n\in\{...,-3,-2,-1,0,1,2,3,...\}~.
\end{gather}
$|\cdot|_p$ denotes the $p$-adic absolute value and $a_{xy}$ is the lowest vertex on the line connecting $x$ and $y$(noted as $\overline{xy}$). Similarly, referring to the right in Fig~\ref{tpandtp2}, $\textrm{T}_{p^2}$ is a tree where each vertex has $p^2+1$ nearest neighboring vertices and its boundary includes $\mathbb{Q}_{p^2}$. As for the dimension of $p$-adic numbers, we assume that $x\in\mathbb{Q}_p$ is dimensionless while $|x|_p$ has the dimension of length. For example, we should write $|x|_p^2=|x^2|_p|1|_p$. To make it simple, $|1|_p$'s are always ignored in this paper, so $|x|_p^2=|x^2|_p$ is also correct. 

\section{Distances on the boundaries of $\textrm{T}_p$ and $\textrm{T}_{p^2}$}\label{sec:disonbdy}

The distance between two vertices on $\textrm{T}_p$ or $\textrm{T}_{p^2}$ can be defined as the number of edges between them
\begin{gather}
d(a,b):=\textrm{number of edges between vertices $a$ and $b$}~.
\end{gather}
It is divergent when $a$ or $b$ go to the boundary. Hence, we need to find another definition of distance for boundary points. In this paper, we consider one kind of distance on the boundary which depends on a selected subgraph(the reference subgraph, noted as $\Omega$). Letting $x,y$ denote two boundary points.

Firstly, we define the distance between line $\overline{xy}$ and $\Omega$ as
\begin{gather}\label{disonbdy}
\log_pd(\overline{xy},\Omega):=\left\{
\begin{aligned}
&\min_{\substack{a\in\overline{xy}\\b\in\Omega}}d(a,b)~,~\textrm{no common edges between $\overline{xy}$ and $\Omega$}~,
\\
&-\frac{1}{2}(\textrm{number of common edges between $\overline{xy}$ and $\Omega$})~,~\textrm{else}~.
\end{aligned}
\right.
\end{gather}
When there are common edges, it can be found that the more common edges they($\overline{xy}$ and $\Omega$) have, the shorter distance there is. Coefficient $1/2$ is chosen because it can be found later that expressions of $d(\overline{xy},\Omega)$ are the same with and without common edges in the case of $\Omega=\overline{0\infty}$ of $\textrm{T}_p$ or $\Omega=\textrm{T}_p$ of $\textrm{T}_{p^2}$.

\begin{figure}
	\centering
	\includegraphics[width=0.8\textwidth]{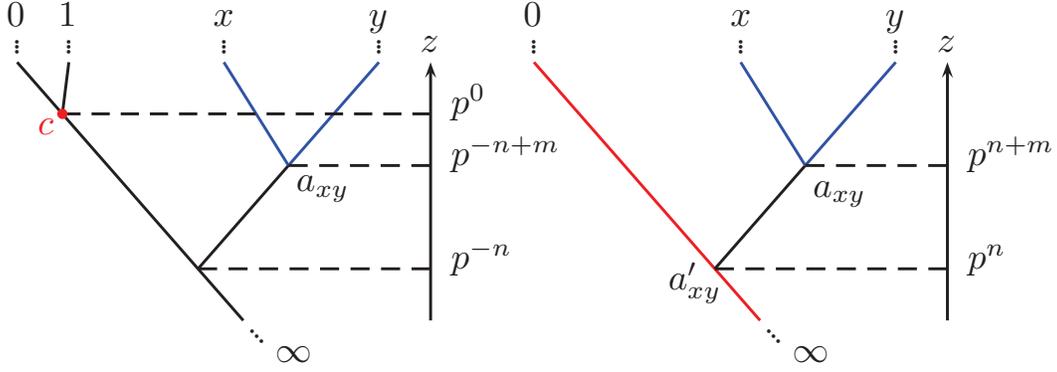}
	\caption{\label{difref}Taking $p=2$ as an example, two different choices of $\Omega$ are considered. Left:~$\Omega=c(0,1,\infty)$; Right:~$\Omega=\overline{0\infty}$. All $\Omega$'s and $\overline{xy}$'s are highlighted in red and blue respectively. Unimportant vertices and edges are not drawn.}
\end{figure}

\begin{figure}
	\centering
	\includegraphics[width=0.8\textwidth]{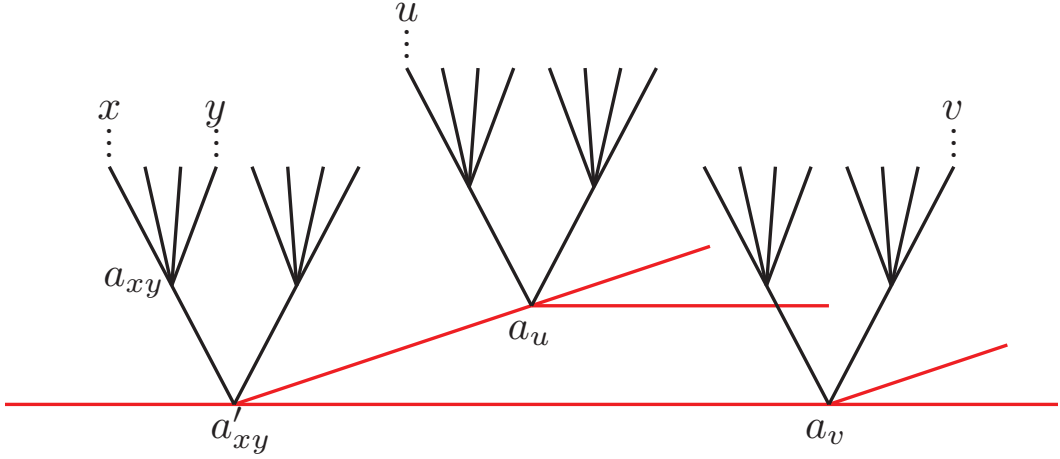}
	\caption{\label{pads2}Taking $p=2$ as an example, $\Omega=\textrm{T}_p$(highlighted in red) is the reference subgraph of $\textrm{T}_{p^2}$. It can be found that $\log_pu_p(x,y)=-2=-2d(a_{xy},a_{xy}')=-2\log_pd(\overline{xy},\Omega)$ and $\log_pu_p(u,v)=2=d(a_u,a_v)=-2\log_pd(\overline{xy},\Omega)$.}
\end{figure}

Secondly, we give three examples of $d(\overline{xy},\Omega)$. Denote the common vertex of lines $\overline{01}$, $\overline{0\infty}$ and $\overline{1\infty}$ as $c(0,1,\infty)$. In the case of $\Omega=c(0,1,\infty)$ of $\textrm{T}_p$, there is no common edges since there is no edge in $\Omega$ at all. The expression of $d(\overline{xy},c)$ writes
\begin{gather}
d(\overline{xy},c)=\left\{
\begin{aligned}
\frac{1}{|x-y|_p}~&,~\textrm{when~}|x|_p~,~|y|_p\leq1~,
\\
\frac{1}{|\frac{1}{x}-\frac{1}{y}|_p}~&,~\textrm{when~}|x|_p~,~|y|_p>1~,
\\
1~&,~\textrm{else}~.
\end{aligned}
\right.
\end{gather}
The proof is simple. For example, referring to the left in Fig~\ref{difref},
when $|x|_p=|y|_p>1$ we have
\begin{gather}
\left.
\begin{aligned}
\log_pd(\overline{xy},c)=\min_{a\in\overline{xy}}d(a,c)=d(a_{xy},c)=&n+m
\\
|\frac{1}{x}-\frac{1}{y}|_p=\frac{|x-y|_p}{|xy|_p}=\frac{|p^{-n+m}|_p}{|p^{-2n}|_p}=&p^{-n-m}
\end{aligned}
\right\}\Rightarrow d(\overline{xy},c)=\frac{1}{|\frac{1}{x}-\frac{1}{y}|_p}~.
\end{gather}
In the case of $\Omega=\overline{0\infty}$ of $\textrm{T}_p$, the expression of $d(\overline{xy},\overline{0\infty})$ writes
\begin{gather}
d(\overline{xy},\overline{0\infty})=\sqrt{\frac{|xy|_p}{|x-y|_p^2}}~.
\end{gather}
The proof is also simple. For example, referring to the right in Fig~\ref{difref}, when there is no common edges we have
\begin{gather}
\left.
\begin{aligned}
\log_pd(\overline{xy},\overline{0\infty})=\min_{\substack{a\in\overline{xy}\\b\in\overline{0\infty}}}d(a,b)=d(a_{xy},a_{xy}')=&m
\\
\frac{|xy|_p}{|x-y|_p^2}=\frac{|p^{2n}|_p}{|p^{n+m}|_p^2}=&p^{2m}
\end{aligned}
\right\}\Rightarrow d(\overline{xy},\overline{0\infty})=\sqrt{\frac{|xy|_p}{|x-y|_p^2}}~.
\end{gather}
The situation is complicated when $\Omega=\textrm{T}_p$ of $\textrm{T}_{p^2}$. Fortunately, it is closely related to the $u_p$ distance of $p\textrm{AdS}$ space(noted as $p\textrm{AdS}_2$ in this paper) which is already studied in~\cite{Gubser:2016guj}. Another useful reference is~\cite{Qu:2018ned}. It is already known that
\begin{gather}
\begin{aligned}
\log_pu_p(x(x_0,x_1),y(y_0,y_1)):=&\log_p\frac{|x_0-y_0,x_1-y_1|_s^2}{|x_0y_0|_p}
\\
=&\left\{
\begin{aligned}
&-2\min_{\substack{a\in\overline{xy}\\b\in\Omega=\textrm{T}_p}}d(a,b)~,~\textrm{no common edges between $\overline{xy}$ and $\Omega=\textrm{T}_p$}~,
\\
&\textrm{number of common edges}~,~\textrm{else}~,
\end{aligned}
\right.
\end{aligned}
\end{gather}
where $|x,y|_s=\max\{|x|_p,|y|_p\}$. Some examples are provided in Fig~\ref{pads2}.
Comparing with \ref{disonbdy}, we have
\begin{gather}
d(\overline{xy},\textrm{T}_p)=\frac{1}{\sqrt{u_p(x,y)}}=\sqrt{\frac{|x_0y_0|_p}{|x_0-y_0,x_1-y_1|_s^2}}~.
\end{gather}

Finally, we can define distances on the boundaries of $\textrm{T}_p$ and $\textrm{T}_{p^2}$. Considering that when $x\to y$ on the boundary, the distance between them should go to zero, we define
\begin{gather}
d_{\Omega}(x,y):=\frac{1}{d(\overline{xy},\Omega)}~,
\end{gather}
which is a distance depending on the reference subgraph $\Omega$. Be aware that $d(a,b)$ is the distance between two vertices and $d_{\Omega}(x,y)$ is the distance between two boundary points. The first two examples are important to this paper, and later we will find that the boundary of $\textrm{T}_p$ becomes $\textrm{EdS}_1$($\textrm{EAdS}_1$) over $\mathbb{Q}_p$ when imposing distance $d_{\Omega=c(0,1,\infty)}$($d_{\Omega=\overline{0\infty}}$). The third example shows that the boundary of $\textrm{T}_{p^2}$ becomes $p\textrm{AdS}_2$ when imposing distance $d_{\Omega=\textrm{T}_p}=\sqrt{u_p}$. 

\section{$\textrm{E(A)dS}_2$ over $\mathbb{Q}_p$ and subspaces}\label{sec:peads}

Referring to the Wick rotation over $\mathbb{R}$
\begin{gather}
x^2\xrightarrow{x\to\sqrt{-1}x} -x^2~,
\\
x\in\mathbb{R}~,~\sqrt{-1}\notin\mathbb{R}~,
\end{gather}
the counterpart over $~\mathbb{Q}_p~$ can be defined as
\begin{gather}
x^2\xrightarrow{x\to\sqrt{\epsilon}x} \epsilon x^2~,
\\
x\in\mathbb{Q}_p~,~\sqrt{\epsilon}\notin\mathbb{Q}_p~.
\end{gather}
$\epsilon$ is not a square of any $p$-adic number. In this paper, we only consider the case of $\epsilon=-1$, and according to~\cite{Vladimirov:1994wi} it demands that
\begin{gather}
p\equiv3\pmod{4}~.
\end{gather}
Several useful properties follow, such as
\begin{gather}
|x^2,y^2|_s=|x^2+y^2|_p~,\label{prop1}
\end{gather}
whose proof can be found in~\cite{Vladimirov:1994wi}.

The continuous $\textrm{EAdS}_2$ space over $\mathbb{Q}_p$($p\textrm{AdS}_2$) is proposed in~\cite{Gubser:2016guj}, where the distance $D(X,Y)$ between $X(x_0',x_1')$ and $Y(y_0',y_1')$ writes
\begin{gather}
p\textrm{AdS}_2:~D^2(X,Y)=u_p(X,Y)=d_{\Omega=\textrm{T}_p}^2(X,Y)=\frac{|x_0'-y_0',x_1'-y_1'|_s^2}{|x_0'y_0'|_p}=\left|\frac{(x_0'-y_0')^2+(x_1'-y_1')^2}{x_0'y_0'}\right|_p~,
\\
x_0',y_0'\in\mathbb{Q}_p^{\times}~,~x_1',y_1'\in\mathbb{Q}_p~.
\end{gather}
$\mathbb{Q}_p^{\times}$ denotes nonzero $p$-adic numbers. The property (\ref{prop1}) is used. With the help of coordinate transformation (\ref{trans2}) and setting $L=1$, we can find the hypersurface equation of $p\textrm{AdS}_2$ and the expression of $D(X,Y)$ using embedding coordinates
\begin{gather}
p\textrm{AdS}_2:~X_0^2+X_1^2-X_2^2=-1~,
\\
D^2(X,Y)=|(X_0-Y_0)^2+(X_1-Y_1)^2-(X_2-Y_2)^2|_p~.
\end{gather}
It is the embedding of $p\textrm{AdS}_2$ into three-dimensional space over $\mathbb{Q}_p$. Although it is better to consider a more general case
\begin{gather}
p\textrm{AdS}_2:~X_0^2+X_1^2-X_2^2=-L^2~,
\end{gather}
we still set $L=1$ for simplicity, and they are related by a scale transformation. Different from the case over $\mathbb{R}$, there is no relation ``$>$'' or ``$<$'' between $p$-adic numbers, and it is meaningless to write down expression ``$x>y$'' or ``$x<y$'' when $x,y\in\mathbb{Q}_p$. So we cannot use constraint $x_0'>0$ or $x_0'<0$ to make $(x_0',x_1')$ coordinate system only cover one half of $p\textrm{AdS}_2$. This problem can be solved in another way. Considering that
\begin{gather}
x>0\Leftrightarrow\textrm{sgn}(x)=1~,~x\in\mathbb{R}^{\times}~,
\end{gather}
where $\mathbb{R}^{\times}$ denotes nonzero real numbers, it is possible to impose the constraint ``$\textrm{sgn}(x_0')=1$'' on $p\textrm{AdS}_2$, since sign functions of $\mathbb{Q}_p^{\times}$ are well defined~\cite{Gubser:2018cha}. In this paper, we do not impose this constraint because $x_0'$ coordinate in the original paper~\cite{Gubser:2016guj} takes value in $\mathbb{Q}_p^{\times}$ freely, containing both cases of $\textrm{sgn}(x_0')=1$ and $\textrm{sgn}(x_0')=-1$.

Referring to the case over $\mathbb{R}$, the $p$-adic version of $\textrm{EdS}_2$ space(noted as $p\textrm{dS}_2$)can be obtained by applying Wick rotation on $p\textrm{AdS}_2$
\begin{gather}
p\textrm{dS}_2:~X_0^2+X_1^2+X_2^2=1~,
\\
D^2(X,Y)=|(X_0-Y_0)^2+(X_1-Y_1)^2+(X_2-Y_2)^2|_p~.
\end{gather}
With the help of coordinate transformations (\ref{trans1}) and setting $L=1$, $D(X,Y)$'s of $p\textrm{(A)dS}_2$ can be collectively written as
\begin{gather}
p\textrm{(A)dS}_2:~D^2(X,Y)=\left|\frac{(x_0-y_0)^2+(x_1-y_1)^2}{(1\pm x_0^2\pm x_1^2)(1\pm y_0^2\pm y_1^2)}\right|_p~,
\end{gather}
where ``$\pm$'' should be replaced by ``$+$'' for $p\textrm{dS}_2$ and ``$-$'' for $p\textrm{AdS}_2$. $|4|_p=1$(when $p\neq2$) is used. It seems difficult to find a graph representation of $p\textrm{dS}_2$ just as $\textrm{T}_{p^2}$ of $p\textrm{AdS}_2$ in Fig~\ref{pads2}, and we cannot solve this problem right now.

Consider subspaces of $p\textrm{(A)dS}_2$
\begin{gather}
p\textrm{(A)dS}_1:~\left\{
\begin{aligned}
X_0^2+X_1^2\pm X_2^2=&\pm1~,
\\
X_1=0~.
\end{aligned}
\right.
\end{gather}
For $p\textrm{dS}_1$, considering that ``$X_1=0\Leftrightarrow x_1=0$'', it can be found in $(x_0,x_1)$ coordinate system that
\begin{gather}
p\textrm{dS}_1:~D(X,Y)=\sqrt{\left|\frac{(x_0-y_0)^2}{(1+x_0^2)(1+y_0^2)}\right|_p}=\left\{
\begin{aligned}
|x_0-y_0|_p~&,~\textrm{when~}|x_0|_p~,~|y_0|_p\leq1
\\
\left|\frac{1}{x_0}-\frac{1}{y_0}\right|_p~&,~\textrm{when~}|x_0|_p~,~|y_0|_p>1
\\
1~&,~\textrm{else}
\end{aligned}
\right\}=d_{\Omega=c(0,1,\infty)}(x_0,y_0)~.
\end{gather}
The property (\ref{prop1}) is used. So $p\textrm{dS}_1$ is the boundary of $\textrm{T}_p$ equipped with distance $d_{\Omega=c(0,1,\infty)}$. For $p\textrm{AdS}_1$, considering that ``$X_1=0\Leftrightarrow x_1'=0$'', it can be found in $(x_0',x_1')$ coordinate system that
\begin{gather}
p\textrm{AdS}_1:~D(X,Y)=\sqrt{\left|\frac{(x_0'-y_0')^2}{x_0'y_0'}\right|_p}=d_{\Omega=\overline{0\infty}}(x_0',y_0')~.
\end{gather}
So $p\textrm{AdS}_1$ is the boundary of $\textrm{T}_p$ equipped with distance $d_{\Omega=\overline{0\infty}}$~.

\section{Summary and discussion}\label{sec:sumanddis}

In this paper, firstly, we define distance on the boundary of $\textrm{T}_p$ or $\textrm{T}_{p^2}$ as $d_{\Omega}(x,y)$ which depends on a subgraph $\Omega$
\begin{gather}
\log_pd_{\Omega}(x,y)=\left\{
\begin{aligned}
&-\min_{\substack{a\in\overline{xy}\\b\in\Omega}}d(a,b)~,~\textrm{no common edges between $\overline{xy}$ and $\Omega$}~,
\\
&\frac{1}{2}(\textrm{number of common edges})~,~\textrm{else}~.
\end{aligned}
\right.
\end{gather}
Secondly, we clarify the Wick rotation($x^2\to-x^2$) over $\mathbb{Q}_p$ which demands that
\begin{gather}
\sqrt{-1}\notin\mathbb{Q}_p\Rightarrow p\equiv3\pmod{4}~.
\end{gather}
With the help of $p\textrm{AdS}_2$ in~\cite{Gubser:2016guj}, we find embedding equations of $p\textrm{(A)dS}_2$
\begin{gather}
X_0^2+X_1^2\pm X_2^2=\pm1~.
\end{gather}
The corresponding distance functions in high-dimensional spaces write
\begin{gather}
D^2(X,Y)=|(X_0-Y_0)^2+(X_1-Y_1)^2\pm(X_2-Y_2)^2|_p~,
\end{gather}
Thirdly, we study $p\textrm{(A)dS}_1$ and compare them with $p\textrm{AdS}_2$. It is found that
\begin{gather}
\begin{aligned}
p\textrm{dS}_1:&~D(X,Y)=d_{\Omega=c(0,1,\infty)}(X,Y)~,
\\
p\textrm{AdS}_1:&~D(X,Y)=d_{\Omega=\overline{0\infty}}(X,Y)~,
\\
p\textrm{AdS}_2:&~D(X,Y)=d_{\Omega=\textrm{T}_p}(X,Y)~.
\end{aligned}
\end{gather}
Hence, $p\textrm{(A)dS}_1$ and $p\textrm{AdS}_2$ can be regarded as boundaries of $\textrm{T}_p$ and $\textrm{T}_{p^2}$ equipped with different $d_{\Omega}$'s. 

There are still many interesting questions needing to be answered. For example, (\romannumeral1)we wonder whether there is a subgraph $\Omega$ of $\textrm{T}_{p^2}$ satisfying
\begin{gather}
p\textrm{dS}_2:~D(X,Y)=d_{\Omega}(X,Y)~;
\end{gather}
(\romannumeral2)with the Wick rotation over $\mathbb{Q}_p$ in hand, we can study the non-Euclidean version of $p\textrm{(A)dS}$ spaces which is not done in this paper; (\romannumeral3)what are the other kinds of distances on the boundary of $\textrm{T}_p$ or $\textrm{T}_{p^2}$ besides those studied in this paper which depend on reference subgraphs; (\romannumeral4)the embedding of $p$-adic version of AdS space has also been studied in other papers such as~\cite{Guilloux_2016,Bhowmick:2018bmn}, but we do not know the relation between our results and theirs yet.

\section*{Acknowledgement}

This work is supported by NSFC Grant No. 11875082.

\bibliography{ref}

\begin{thebibliography}{10}

\bibitem{Volovich:1987wu}
Igor~V. Volovich.
\newblock Number theory as the ultimate physical theory.
\newblock {\em P-Adic Numb. Ultrametr. Anal. Appl.}, 2:77--87, 2010.
\newblock doi: {10.1134/S2070046610010061}.

\bibitem{Volovich_1987}
Igor~V. Volovich.
\newblock p-adic string.
\newblock {\em Classical and Quantum Gravity}, 4(4):L83--L87, jul 1987.
\newblock doi: {10.1088/0264-9381/4/4/003}.

\bibitem{Freund:1987kt}
Peter G.~O. Freund and Mark Olson.
\newblock {NONARCHIMEDEAN STRINGS}.
\newblock {\em Phys. Lett. B}, 199:186--190, 1987.
\newblock doi: {10.1016/0370-2693(87)91356-6}.

\bibitem{FREUND1987191}
Peter~G.O. Freund and Edward Witten.
\newblock Adelic string amplitudes.
\newblock {\em Physics Letters B}, 199(2):191--194, 1987.
\newblock doi: {10.1016/0370-2693(87)91357-8}.

\bibitem{Zabrodin:1988ep}
Anton~V. Zabrodin.
\newblock {Nonarchimedean Strings and Bruhat-tits Trees}.
\newblock {\em Commun. Math. Phys.}, 123:463, 1989.
\newblock doi: {10.1007/BF01238811}.

\bibitem{Vladimirov:1988bd}
V.~S. Vladimirov and I.~V. Volovich.
\newblock {P-ADIC QUANTUM MECHANICS}.
\newblock {\em Sov. Phys. Dokl.}, 33:669--670, 1988.
\newblock doi: {10.1007/BF01218590}.

\bibitem{Smirnov:1992sr}
Vladimir~A. Smirnov.
\newblock {Calculation of general p-adic Feynman amplitude}.
\newblock {\em Commun. Math. Phys.}, 149:623--636, 1992.
\newblock doi: {10.1007/BF02096946}.

\bibitem{Gubser:2017vgc}
Steven~S. Gubser, Christian Jepsen, Sarthak Parikh, and Brian Trundy.
\newblock {O(N) and O(N) and O(N)}.
\newblock {\em JHEP}, 11:107, 2017.
\newblock doi: {10.1007/JHEP11(2017)107}.

\bibitem{Vladimirov:1994wi}
V.~S. Vladimirov, I.~V. Volovich, and E.~I. Zelenov.
\newblock {\em {p-adic analysis and mathematical physics}}, volume~1.
\newblock WORLD SCIENTIFIC, 1994.
\newblock doi: {10.1142/1581}.

\bibitem{Heydeman:2016ldy}
Matthew Heydeman, Matilde Marcolli, Ingmar Saberi, and Bogdan Stoica.
\newblock {Tensor networks, $p$-adic fields, and algebraic curves: arithmetic
  and the AdS$_3$/CFT$_2$ correspondence}.
\newblock {\em Adv. Theor. Math. Phys.}, 22:93--176, 2018.
\newblock doi: {10.4310/ATMP.2018.v22.n1.a4}.

\bibitem{Gubser:2016guj}
Steven~S. Gubser, Johannes Knaute, Sarthak Parikh, Andreas Samberg, and Przemek
  Witaszczyk.
\newblock {$p$-adic AdS/CFT}.
\newblock {\em Commun. Math. Phys.}, 352(3):1019--1059, 2017.
\newblock doi: {10.1007/s00220-016-2813-6}.

\bibitem{Maldacena:1997re}
Juan~Martin Maldacena.
\newblock {The Large N limit of superconformal field theories and
  supergravity}.
\newblock {\em Adv. Theor. Math. Phys.}, 2:231--252, 1998.
\newblock doi: {10.1023/A:1026654312961}.

\bibitem{Gubser:1998bc}
Steven~S. Gubser, Igor~R. Klebanov, and Alexander~M. Polyakov.
\newblock {Gauge theory correlators from noncritical string theory}.
\newblock {\em Phys. Lett. B}, 428:105--114, 1998.
\newblock doi: {10.1016/S0370-2693(98)00377-3}.

\bibitem{Witten:1998qj}
Edward Witten.
\newblock {Anti-de Sitter space and holography}.
\newblock {\em Adv. Theor. Math. Phys.}, 2:253--291, 1998.
\newblock doi: {10.4310/ATMP.1998.v2.n2.a2}.

\bibitem{Gubser:2016htz}
Steven~S. Gubser, Matthew Heydeman, Christian Jepsen, Matilde Marcolli, Sarthak
  Parikh, Ingmar Saberi, Bogdan Stoica, and Brian Trundy.
\newblock {Edge length dynamics on graphs with applications to $p$-adic
  AdS/CFT}.
\newblock {\em JHEP}, 06:157, 2017.
\newblock doi: {10.1007/JHEP06(2017)157}.

\bibitem{Bhattacharyya:2017aly}
Arpan Bhattacharyya, Ling-Yan Hung, Yang Lei, and Wei Li.
\newblock {Tensor network and ($p$-adic) AdS/CFT}.
\newblock {\em JHEP}, 01:139, 2018.
\newblock doi: {10.1007/JHEP01(2018)139}.

\bibitem{Qu:2018ned}
Feng Qu and Yi-hong Gao.
\newblock {Scalar fields on $p$AdS}.
\newblock {\em Phys. Lett. B}, 786:165--170, 2018.
\newblock doi: {10.1016/j.physletb.2018.09.043}.

\bibitem{Gubser:2018cha}
Steven~S. Gubser, Christian Jepsen, and Brian Trundy.
\newblock {Spin in $p$-adic AdS/CFT}.
\newblock {\em J. Phys. A}, 52(14):144004, 2019.
\newblock doi: {10.1088/1751-8121/ab0757}.

\bibitem{Ebert:2019src}
Stephen Ebert, Hao-Yu Sun, and Meng-Yang Zhang.
\newblock {Probing holography in $p$-adic CFT}.
\newblock 11 2019.
\newblock arXiv: {1911.06313}.

\bibitem{Huang:2019nog}
An~Huang, Bogdan Stoica, and Shing-Tung Yau.
\newblock {General relativity from $p$-adic strings}.
\newblock 1 2019.
\newblock arXiv: {1901.02013}.

\bibitem{Huang:2020qik}
An~Huang, Bogdan Stoica, Xuyang Xia, and Xiao Zhong.
\newblock {Bounds on the Ricci curvature and solutions to the Einstein
  equations for weighted graphs}.
\newblock 6 2020.
\newblock arXiv: {2006.06716}.

\bibitem{Guilloux_2016}
Antonin Guilloux.
\newblock Yet another $p$-adic hyperbolic disc: Hilbert distance for $p$-adic
  fields.
\newblock {\em Groups, Geometry, and Dynamics}, 10(1):9–43, 2016.
\newblock doi: {10.4171/ggd/341}.

\bibitem{Bhowmick:2018bmn}
Samrat Bhowmick and Koushik Ray.
\newblock {Holography on local fields via Radon Transform}.
\newblock {\em JHEP}, 09:126, 2018.
\newblock doi: {10.1007/JHEP09(2018)126}.

\bibitem{Stoica:2018zmi}
Bogdan Stoica.
\newblock {Building Archimedean Space}.
\newblock 9 2018.
\newblock arXiv: {1809.01165}.

\end{thebibliography}

\end{document}